\documentclass[twocolumn,prb,aps,citeautoscript,showpacs,footinbib,amsmath,amssymb,superscriptaddress,longbibliography]{revtex4-1}
\usepackage{array}
\usepackage{graphicx}
\usepackage{dcolumn}
\usepackage{color}
\usepackage{bm}
\usepackage{natbib}
\usepackage{hyperref}
\usepackage{amsmath,amssymb,amsfonts,bbold}
\usepackage[normalem]{ulem}
\usepackage{float}

\usepackage{tikz}

\newcommand{\beq}{\begin{equation}}
\newcommand{\eeq}{\end{equation}}
\newcommand{\beqa}{\begin{eqnarray}}
\newcommand{\eeqa}{\end{eqnarray}}

\begin{document}
\title{Probing spin-dependent charge transport at single-nanometer length scales}

\author{Patrick H\"{a}rtl}
	\email[corresponding author: ]{patrick.haertl@physik.uni-wuerzburg.de}
	\address{Physikalisches Institut, Experimentelle Physik II, 
	Universit\"{a}t W\"{u}rzburg, Am Hubland, 97074 W\"{u}rzburg, Germany}
\author{Markus Leisegang}
	\affiliation{Physikalisches Institut, Experimentelle Physik II, 
	Universit\"{a}t W\"{u}rzburg, Am Hubland, 97074 W\"{u}rzburg, Germany} 
\author{Jens K\"{u}gel}
	\affiliation{Physikalisches Institut, Experimentelle Physik II, 
	Universit\"{a}t W\"{u}rzburg, Am Hubland, 97074 W\"{u}rzburg, Germany} 
\author{Matthias Bode} 
	\address{Physikalisches Institut, Experimentelle Physik II, 
	Universit\"{a}t W\"{u}rzburg, Am Hubland, 97074 W\"{u}rzburg, Germany}	
	\address{Wilhelm Conrad R\"{o}ntgen-Center for Complex Material Systems (RCCM), 
	Universit\"{a}t W\"{u}rzburg, Am Hubland, 97074 W\"{u}rzburg, Germany}       


\date{\today}

\begin{abstract}
The coherent transport of charge and spin is one key requirement of future devices for quantum computing and communication. 
Scattering at defects or impurities may seriously reduce the coherence of quantum-mechanical states, thereby affecting device functionality. 
While numerous methods exist to experimentally assess charge transport, 
the real-space detection of a material's spin transport properties with nanometer resolution remains a challenge.  
Here we report on a novel approach which utilizes a combination of spin-polarized scanning tunneling microscopy (SP-STM) 
and the recently introduced molecular nanoprobe (MONA) technique. 
It relies on the local injection of spin-polarized charge carriers from a magnetic STM tip and their detection 
by a single surface-deposited phthalocyanine molecule via reversible electron-induced tautomerization events.  
Based on the particular electronic structure of the Rashba alloy BiAg$_2$ which is governed by a spin--momentum-locked surface state,
we proof that the current direction inverses as the tip magnetization is reversed by an external field.  
In a proof-of-principle experiment we apply SP-MONA to investigate how a single Gd cluster 
influences the spin-dependent charge transport of the Rashba surface alloy. 
\end{abstract}


\maketitle


\section*{Introduction}
\label{sect:Introduction}
\vspace{-0.3cm}

The progressing miniaturization of electronics components in integrated circuits 
has reached a point where single defects \cite{Miroshnichenko2010,Waltl2020} and the coherent superposition 
of quantum-mechanical states \cite{Makhlin2001,Gabelli2006,Miroshnichenko2010} have to be considered.  
In fact, future technologies may fundamentally rely on nonlocal phase-coherent charge transfer processes, 
thereby enabling novel device concepts which materialize the enormous gain promised by quantum computation and communication,  
e.g., by utilizing Josephson tunneling junctions \cite{Makhlin2001} or zero-energy Majorana bound states \cite{Fu2010}.  

Particularly fascinating are strategies where the conventional manipulation of charge is replaced by the manipulation of the electron spin.  
For a long time, the concept of spintronics relied on the combination 
of non-magnetic semiconductors with magnetic polarizers \cite{Wolf2001,Dieny2020}. 
However, the injection of spin-polarized charge carriers across material interfaces remained a serious challenge \cite{Ohno1999}. 
In this context, the spin--momentum-locking \cite{Hasan2010,Kohda2019,Moore2010,Fu2007,Zhang2009,Hsieh2009} 
of Rashba-split surface or interface states \cite{Bychkov1984,Koo2020} 
or topologically protected boundary states \cite{Konig2007,Sessi2016,Jung2021} 
represents a formidable opportunity to overcome these limitations.  
In fact, the discovery of Aharonov-Bohm oscillations in topological insulators \cite{Peng2010} 
or the observation of Datta-Das oscillation in the ballistic intrinsic spin Hall effect \cite{Choi2015}
clearly demonstrate that the coherent propagation of quantum-mechanical electronic states 
is a viable approach towards future spintronic devices. 

In spite of the high expectations in the combination of spin--momentum-locking and spintronics 
our capabilities in detecting the spatial distribution of spin currents are quite limited.  
The existence of the edge channels has been demonstrated by imaging the current-induced magnetic fields in HgTe quantum wells
by means of SQUID microscopy with $\mu$m resolution \cite{Nowack2013}, but these data lack intrinsic spin sensitivity.  
Optical Kerr imaging methods are able to visualize the spin transport in lateral ferromagnet/semiconductor structures \cite{Crooker2005}, 
but their lateral resolution is limited by the wave length of light.  
Shorter transport distances can be probed by lithographically prepared Hall bars, but the pre-defined electrode configuration 
cannot be changed any more and material damage may occur during processing \cite{Matsuo2012}.
Multi-probe scanning tunneling microscopy (STM) setups offer a much higher spatial resolution 
with inter-tip distances down to about 30\,nm \cite{KHM2005,Miccoli2015,Yang2016,Leis2020,Leis2021,Leis2022}, 
but have not yet been successfully applied with spin-sensitive magnetic tips. 

Recently, we developed the molecular nanoprobe (MONA) technique which is capable 
of detecting ballistic charge transport on length scales down to a few nanometers.  
In this technique, charge carriers locally injected by an STM tip propagate across the surface  
and are detected by a single molecule via a rever\-sible electron-induced switching process, 
such as a tau\-to\-me\-ri\-za\-tion \cite{Kuegel2017a}.  
Charge transport in surface states \cite{Kuegel2017,Christ2022,Leisegang2023}, 
anisotropic transport on fcc(110) surfaces \cite{Leisegang2021}, 
and the damping and amplification by coherent superposition of quantum-mechanical waves 
in engineered atomic-scale structures has been experimentally demonstrated \cite{Leisegang2018}.  

\begin{figure}[t]
	\centering
	\includegraphics[width=\linewidth]{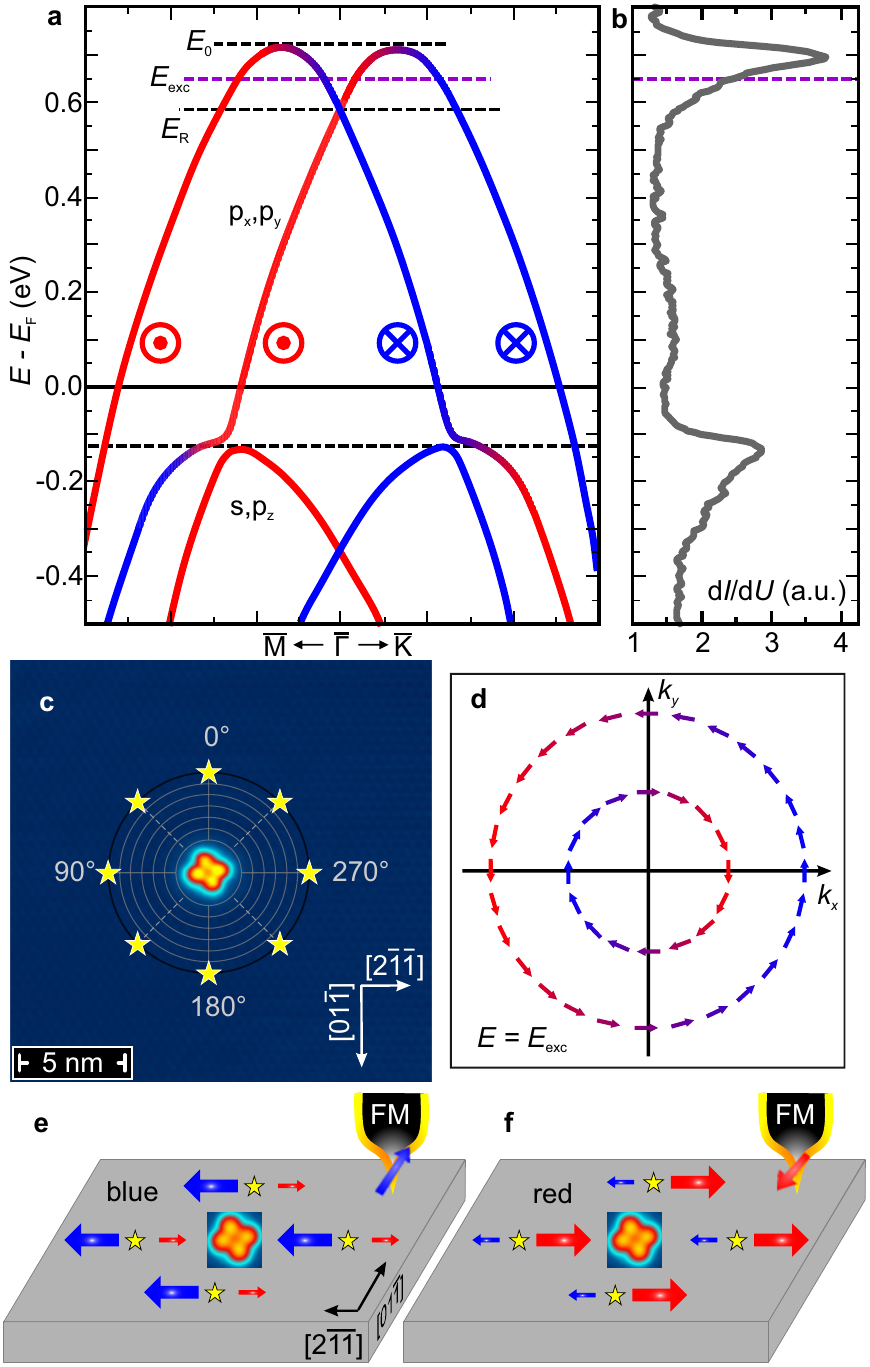}
			\caption{\textbf{The BiAg$_2$ Rashba-split surface state.} 
			\textbf{a}, Schematic representation of the Rashba-split bands with blue and red spin type 
			pointing in and out of the drawing plane, adapted from Ref.\,\onlinecite{Bihlmayer2007}. 
			\textbf{b}, d$I$/d$U$ point spectrum of the BiAg$_2$ surface with peaks indicating the onsets 
			of two downwards dispersing Rashba-split surface band at $E_{1}\approx -130$\,meV and $E_{2}\approx 700$\,meV. 
			\textbf{c}, Constant-energy cut taken at the purple dashed line in (a) with two spin--momentum-locked rings.
			\textbf{d}, STM image of a single HPc molecule on the BiAg$_2$ surface alloy. 
			For MONA measurements, charge carriers are injected at equi-angular positions on a circle around the molecule, 
			as marked by yellow stars. STM parameters: $U_{\text{scan}} = 200$\,mV, $I_{\text{scan}} = 100$\,pA.
			\textbf{e,f}, Schematic drawings for the expected directional transport of a spin-polarized current 
			injected by an SP tip in the unoccupied surface state at the four yellow stars.}
		\label{Fig:1}
\end{figure}

In this study, we report on the development and application of spin-polarized (SP)-MONA.  
The capability of investigating ballistic transport properties of spin-polarized charge carriers in real-space on length scales 
of a few nanometers is demonstrated by utilizing spin--momentum-locked Rashba-split bands of the BiAg$_2$ surface alloy. 
As shown in Fig.\,\ref{Fig:1}(a), BiAg$_2$ features two downwards dispersing surface states 
within the $L$-projected bulk band gap, an occupied $s, p_z$-like band and a partially unoccupied $p_x, p_y$-derived band. 
Both bands exhibit a giant Rashba splitting of $E_0 - E_{\text{R}} \approx 200$\,meV \cite{Ast2007,Ast2007a,Bihlmayer2007,Bentmann2009}.
The tunneling spectrum of the BiAg$_2$ surface presented in Fig.\,\ref{Fig:1}(b) shows two peaks 
which indicate the onset energies $E_0$ of the Rashba-split surface states. 
Throughout the entire study, experiments will be performed at an energy $E = E_{\rm exc} = 650$\,meV, 
marked by a purple dashed line in Fig.\,\ref{Fig:1}(a) between $E_0$ and $E_{\rm R}$. 

The unoccupied $p_x, p_y$-derived band exhibits an unconventional spin polarization \cite{ElKareh2014},
characterized by a reversal at the upper onset of the band, $E_0$,
as schematically represented by a transition from red to blue color in Fig.\,\ref{Fig:1}(a). 
This unusual Rashba splitting leads to a spin-dependent propagation of charge carriers injected in the unoccupied bands, 
which will be discussed for states with $k_y = 0$ without limiting the generality of our considerations.
While the electrons carrying a blue-colored spin ($\otimes$) 
move with a negative group velocity $v_{\otimes}^{\rm g} = \nabla_{k} E < 0$, 
electrons with a red-colored spin ($\odot$) propagate in the opposite direction, $v_{\odot}^{\rm g} = \nabla_{k} E > 0$. 
As a consequence, we expect a striking asymmetry of the charge carrier propagation in real space, 
with $\otimes$-electrons propagating to the left and $\odot$-electrons moving to the right.

To analyze this asymmetric propagation with the MONA technique, we manipulated a single phthalocyanine (H$_2$Pc) to a defect-free area 
and subsequently deprotonated it to a detector molecule HPc, see Fig.\,\ref{Fig:1}(d). 
Yellow stars mark the locations where charge carriers are injected from the STM tip directly into the substrate.
(see \hyperref[sec:methods]{\emph{Methods}} for details). 
The charge carrier-induced tautomerization of HPc serves as a measure for transport, presented as normalized electron yield $\eta$ in the following.

\begin{figure*}[t]
	\centering
	\includegraphics[width=0.8\linewidth]{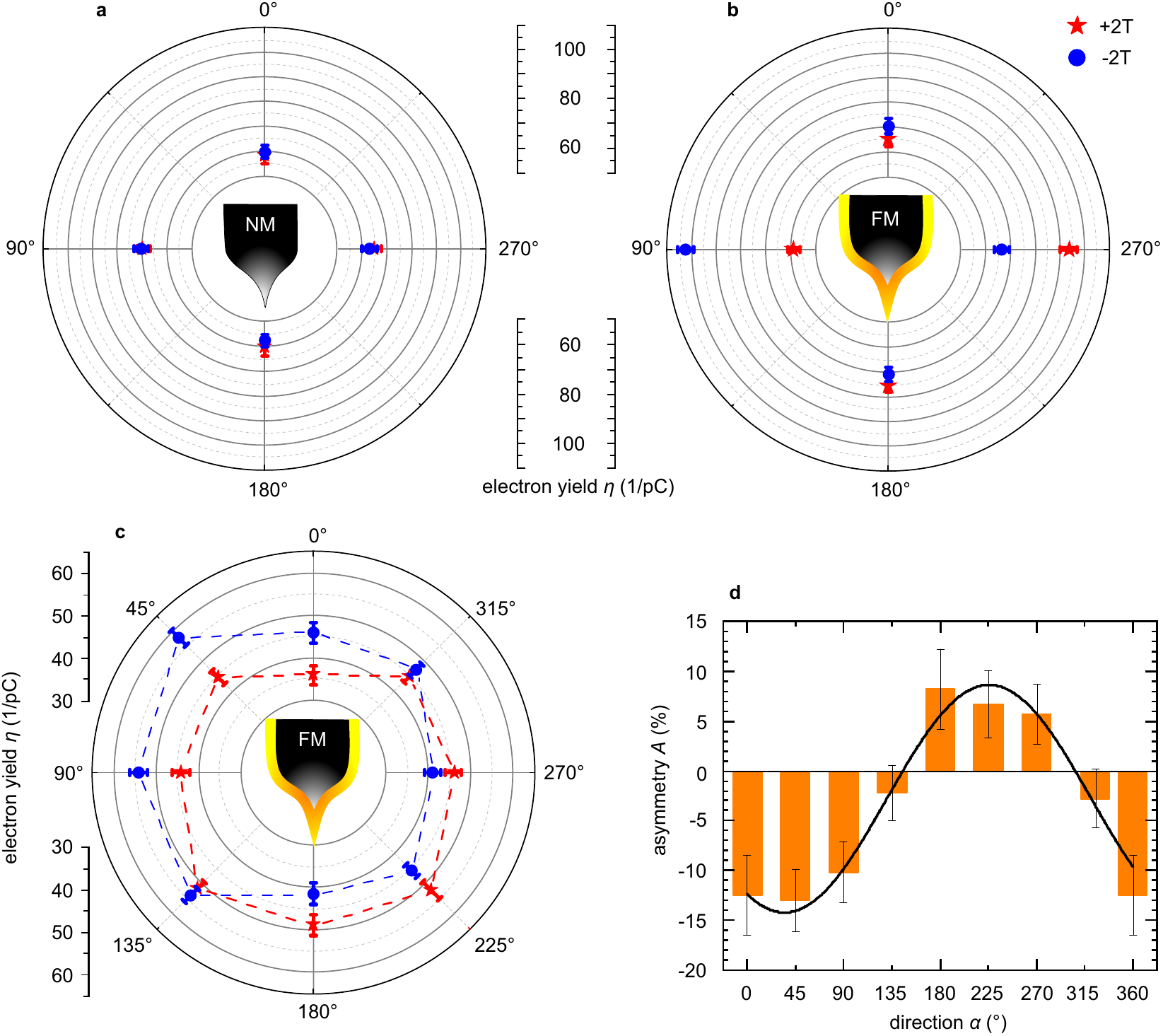}
		\caption{\textbf{SP-MONA results.} \textbf{a}, MONA measurements with a non-magnetic (NM) tungsten tip. 
			The blue/red symbols in the polar plot represent the normalized electron yields $\eta$ 
			at four injection points taken after magnetic field ramps to $\pm 2$\,T and measured in remanence. 
			\textbf{b}, Same measurements as presented in (a) for a ferromagnetic (FM) Gd-coated tip. 
			\textbf{c}, Electron yield $\eta$ taken at eight injection points with a different Gd tip 
			in $45^{\circ}$ steps for both field directions in remanence. The dashed lines serve as a guide to the eye. 
			\textbf{d}, Histogram of the calculated asymmetry $A$ between both field sweeps (orange bars) of (\textbf{c}), 
			indicating a cosine behavior (fit, black line).
			MONA parameters:  Injection time $t_{\text{exc}} = 2.0$\,s, $I_{\text{exc}} = 1.5$\,nA, $E_{\text{exc}} = 650$\,mV. }
		\label{Fig:3}
\end{figure*}

As sketched in Fig.\,\ref{Fig:1}(c), the constant-energy cut at $E_{\rm exc}$ is governed by spin--momentum-locking, 
i.e., spins which are oriented perpendicular to the respective wave vector. 
Charge carriers with such an in-plane spin can be induced from a magnetically coated STM tip in the Rashba bands. 
The resulting asymmetry is expected to be strongest in the direction where the tip magnetization 
is colinear with the spin of the Rashba bands \cite{Meservey1994,Bode2003}.
For electrons with $k_y = 0$ this is the case for a tip magnetized along the 
in-plane $[01\bar{1}]$ direction of BiAg$_2$.  
As sketched in Fig.\,\ref{Fig:1}(e,f), this should lead to the injection of blue $\otimes$-electrons with a negative group velocity, 
resulting in a high (low) transport towards the molecule at $\alpha = 270^{\circ}$ ($\alpha = 90^{\circ}$), i.e. an electron yield $\eta_{270}^{\text{blue}} > \eta_{90}^{\text{blue}}$.
Inverting the in-plane tip magnetization along the $\left[0 \bar{1} 1 \right]$ direction, 
see Fig.\,\ref{Fig:1}(f), would result in the injection of $\odot$-electrons with a positive group velocity. 
As a consequence, the preferred direction of charge transport would also invert, 
i.e.\ we expect $\eta_{90}^{\text{red}} > \eta_{270}^{\text{red}}$. 
To quantify the spin polarization of charge transport when reversing the tip magnetization, 
the asymmetry $A_{\alpha}$ of the electron yields $\eta$ at a given angle $\alpha$ can be calculated as 
$A_{\alpha} = \left(\eta_{\alpha}^{+2\,\text{T}}-\eta_{\alpha}^{-2\,\text{T}}\right)
			/\left(\eta_{\alpha}^{+2\,\text{T}}+\eta_{\alpha}^{-2\,\text{T}} \right)$.
In contrast to a SP tip, the spin-averaged signal of a non-magnetic tip should result in a vanishing asymmetry $A_{\alpha}$.

\section*{Results}
\label{sect:Results}
\vspace{-0.3cm}
In Fig.~\ref{Fig:3} the results of measurements performed (a) with a non-magnetic W tip 
and (b) a Gd-coated magnetic tip are presented in polar coordinates. 
Each tip was treated in an external magnetic field of $\pm 2$\,T (red stars and blue circles, respectively) 
before the data were acquired in remanence ($0$\,T).
Charge carriers were injected with MONA parameters of $E_{\text{exc}} = 650$\,meV, $t_{\text{exc}} = 2.0$\,s, 
$I_{\text{exc}} = 1.0$\,nA at a distance of $d = 4.0$\,nm from the molecule under four different angles. 
The data for a non-magnetic W tip, Fig.\,\ref{Fig:3}(a), show an electron yield $\eta$ 
which, within error bars, is independent of the magnetic history of the tip. 
This can be quantified by an asymmetry $A_{\alpha}^{\text{NM}} < (2 \pm 3)$\%.
The small anisotropy of $\eta$ between $0^{\circ}$/$180^{\circ}$ and $90^{\circ}$/$270^{\circ}$ results from
the anisotropic coupling of the molecule to the substrate, as discussed in detail recently \cite{Leisegang2023}
and quantified in the Suppl.\ Sects.~2 and 3.

\begin{figure*}[t]
	\centering
	\includegraphics[width=0.8\linewidth]{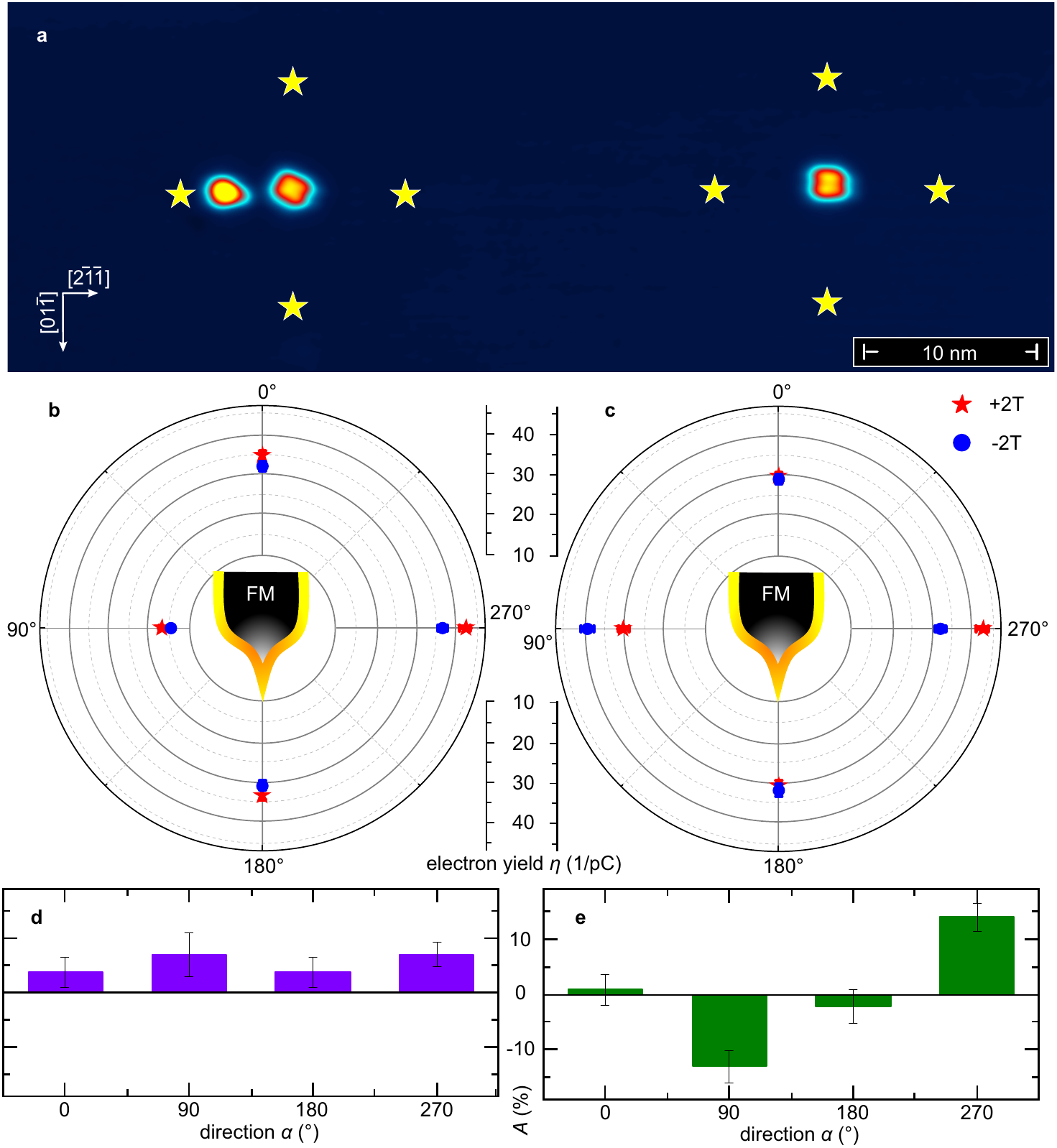}
		\caption{\textbf{Scattering at a Gd-cluster.} 
			\textbf{a}, Topographic STM image of two HPc molecules on a BiAg$_2$ surface, 
			separated by approximately 30\,nm along the $\left[2\bar{1}\bar{1}\right]$ direction. 
			A Gd cluster which serves as a scattering center was deliberately deposited about 4\,nm left of the left HPc molecule.  
			The right molecule is used as a reference system without an additional scatterer. 
			Yellow stars represent the injection points for SP-MONA measurements.
			STM parameters: $U_{\text{scan}} = 200$\,mV, $I_{\text{scan}} = 100$\,pA.
			\textbf{b}, Polar plot of the electron yield $\eta$ for the left molecule with the nearby Gd cluster, 
			measured in remanence after the application of a $\pm 2$\,T field, see red/blue data points, respectively. 
			\textbf{c}, Same as (\textbf{b}) but for the right reference system. 
			\textbf{d,e}, Asymmetries $A$ of the tip magnetization direction-dependent electron yield of panels (\textbf{b,c}). 
			MONA parameters for both molecules:  $t_{\text{exc}} = 2.5$\,s, $I_{\text{exc}} = 2.0$\,nA, $E_{\text{exc}} = 650$\,meV.}
		\label{Fig:4}
\end{figure*}
In contrast, the data presented in Fig.\,\ref{Fig:3}(b) for charge carriers injected from 
a magnetically coated Gd tip reveal a striking difference between the $\pm 2$\,T sweeps. 
While the red and blue data points at $0^{\circ}$ and $180^{\circ}$ 
coincide within the error bars ($A_{0}^{\text{FM}}=(4 \pm 3)$\%, $A_{180}^{\text{FM}}=(3 \pm 3)$\%), 
a significant deviation can be observed at $90^{\circ}$ and $270^{\circ}$. 
The tip treatment at $-2$\,T results in a high (low) electron yield at $90^{\circ}$ ($270^{\circ}$), 
which inverts 
upon a treatment at $+2$\,T. 
Quantitative analysis results in $A_{90}^{\text{FM}} = (27 \pm 3)$\% and $A_{270}^{\text{FM}}=(17 \pm 3)$\%. 

These data are in line with our hypothesis of Fig.~\ref{Fig:1}(e,f).
Indeed, post-characterization of the specific Gd-coated tip used for the experiments of Fig.\,\ref{Fig:3}(b) 
on a test sample with Fe/W(110) monolayer islands confirms a significant in-plane polarization 
along the $0^{\circ}$--$180^{\circ}$ direction which can be inverted upon a field sweep at $\pm 2$\,T, see Suppl.\ Sect.~6. 
Already at this point we can conclude, that the absence of a significant asymmetry for a non-magnetic tip 
in combination with the strong asymmetry observed for the magnetically Gd-coated tip 
proves that SP-MONA allows to detect spin-dependent transport 
in the spin--momentum-locked Rashba-split surface state of the BiAg$_2$ alloy.

To further substantiate this claim, we conducted MONA measurements 
at eight different angles ($\Delta \alpha=45^{\circ}$) with a macroscopically different Gd-coated tip. 
Charge carriers were injected at a distance $d = 4.5$\,nm from the detector molecule.  
In Fig.\,\,\ref{Fig:3}(c) the results measured in remanence after a tip treatment at $\pm2$\,T are shown in a polar plot.
Along the $135^{\circ}$--$315^{\circ}$ direction, the data points for $+2$\,T and $-2$\,T coincide within the error bars, 
whereas a significant difference can be observed for the other six angles. 
The quantitative analysis reveals a cosine-like behavior of the asymmetry $A$, as presented in Fig.\,\ref{Fig:3}(d), 
which can be fitted by $A(\alpha) = O + b \cos(\alpha-\alpha_0)$. 
Hereby, $\alpha_0 = (34.6 \pm 0.1)^{\circ}$ represents the direction with the largest asymmetry, $A_0 = (-14.2 \pm 0.1)$\%. 
We speculate that the offset $O = (-2.8 \pm 0.2)$\% is caused by an imperfect inversion of the tip magnetization during the field sweep,  
resulting in slightly different in-plane projections in remanence.

The experimental data presented so far were obtained on perfect surfaces 
to utilize the well-known spin--momentum-locked electronic structure of the Rashba surface alloy BiAg$_2$.  
The strength of SP-MONA, however, lies in the analysis of imperfect surfaces where charge and spin transport 
is affected by, e.g., the presence of vacancies, interstitials, domain boundaries, or adatoms.  
To demonstrate the capability of SP-MONA, 
we conducted transport measurements across a magnetic cluster deposited on the BiAg$_2$ surface. 
The carefully designed setup is presented in the topographic STM image of Fig.\,\ref{Fig:4}(a). 
It consists of two HPc molecule which are placed at a distance of about 30\,nm on a defect-free region of the BiAg$_2$ surface alloy.  
A Gd cluster was deliberately deposited from the STM tip at a distance of $\approx 4$\,nm from the left molecule. 
This molecule will allow for the spin-dependent detection of charge carrier transport 
injected at the four surrounding injection points (yellow stars, $d = 6.5$\,nm) under the influence of the Gd cluster.  
In contrast, the right pristine HPc molecule is far from any defect or impurity and serves as a reference system. 

Figure~\ref{Fig:4}(b,c) depicts the averaged and normalized electron yields at remanence for the two MONA setups in (a).
While the plot in Fig.\,\ref{Fig:4}(c) with its pronounced asymmetry along the $90^{\circ}$--$270^{\circ}$ direction is in perfect agreement 
with the observation of spin-dependent charge carrier transport in a Rashba-split surface state reported in Fig.\,\ref{Fig:3}(b), 
a strong influence of the cluster can be observed in Fig.\,\ref{Fig:4}(b). 
The overall electron yield across the cluster (data point at $90^\circ$) is significantly reduced, 
in-line with inelastic scattering events which would result in reduced ballistic transport between the injection point and the detector molecule. 
In contrast, the electron yield $\eta$ measured for the other directions are either equivalent or reveal a higher $\eta$ as compared to the right setup.
This may be caused by the constructive superposition of quantum-mechanical states, 
as previously observed in similar experiments \cite{Leisegang2018}.

The calculated asymmetries allow for a discussion of the spin-dependent effects.  
Indeed, the data of the right reference setup, see Fig.\,\ref{Fig:4}(e), are consistent with expected cosine-like behavior. 
In contrast, an overall positive and partially reduced asymmetry is observed under the influence of the Gd cluster, see Fig.\,\ref{Fig:4}(d). 
Since both measurements were conducted with the very same tip, the changes clearly result 
from the presence of the Gd cluster, possibly caused by spin-flip scattering events.   
Especially the reversal at $\alpha = 90^{\circ}$ from $A_{90} = (-14 \pm 3)$\% for the clean surface 
to $A_{90} = (7 \pm 4)$\% across the Gd cluster is quite surprising.  
Since the cluster exhibits an apparent height $h_{\rm Gd} = (173 \pm 5)$\,pm and a diameter $d_{\rm Gd} = (1.16 \pm 0.13)$\,nm only, 
and since Gd with its rather spherical charge distribution usually exhibits a relatively low magneto-crystalline anisotropy, 
we would expect that the cluster is superparamagnetic even at the measurement temperature of about 5\,K. 
The frequent thermally induced magnetization reversals should reduce 
rather than invert the spin polarization of the ballistic current across the cluster.  
We speculate that the interaction of the cluster with the strongly spin--orbit-coupled substrate induces a strong anisotropy 
which---in combination with the magnetic field applied for tip magnetization reversal---is sufficient to induce a remanent magnetization.
However, due to the unknown geometric and magnetic properties of the cluster, precludes a profound analysis.

\section*{Conclusions}
\label{subsec:Conclusion}
\vspace{-0.35cm}

Our study shows that SP-MONA is unique experimental method which allows to probe 
ballistic charge transport properties at previously inaccessible length scales.  
As a STM-derived technique the transport data can directly be correlated to topographic data, 
thereby allowing an assessment how crystallographic imperfections at surface or interfaces affect spin transport. 
While the Rashba-split surface state of the BiAg$_2$ surface provided an ideal testbed to demonstrate the general capabilities of SP-MONA, topological insulators (TIs) \cite{Peng2010} or two-dimensional (2D) materials like graphene \cite{Berger2006} 
or transition metal chalcogenides \cite{Shen2022} will be highly interesting materials for future experiments.

\section*{Acknowledgments}
\label{sect:Acknowledgments}
\vspace{-0.35cm}
This work was supported by the DFG through SFB 1170 (project A02).  
We also acknowledge financial support by the Deutsche Forschungsgemeinschaft (DFG, German Research Foundation) 
under Germany's Excellence Strategy through W{\"u}rzburg-Dresden Cluster of Excellence 
on Complexity and Topology in Quantum Matter -- ct.qmat (EXC 2147, project-id 390858490).

\section*{Contributions}
\label{sect:Contributions}
\vspace{-0.35cm}
P.H. and M.L. performed the experiments and analyzed the resulting data with input from M.B. The experiments were conceived and designed by all authors. Experimental procedures and analysis tools were established by J.K. and M.L. and conducted by P.H. and M.L. P.H., M.L. and M.B. wrote the manuscript with input from J.K.

\section*{Data availability}
\label{sect:Data availability}
\vspace{-0.35cm}
The data that support the findings of this study are available from the corresponding authors upon reasonable request.

\section*{Methods}
\label{sec:methods}
\vspace{-0.35cm}

\subsection*{Experimental setup}
\label{subsec:setup}
\vspace{-0.35cm}
The results were obtained in a two-chamber ultra-high vacuum (UHV) system (base pressure ${p\leq 5\times 10^{-11}}$\,mbar). 
STM measurements were carried out in the constant-current mode with a home-built low-temperature system 
with a base temperature of ${T_{\text{STM}}\approx 4.5}$\,K, with the bias voltage applied to the sample. 
A magnetic field oriented perpendicular to the surface plane with ${\lvert\mu_{0}H\rvert\leq 3}$\,T 
can be generated by a superconducting split-coil magnet.
\vspace{-0.3cm}

\subsection*{Sample preparation}
\label{subsec:sample_prep}
\vspace{-0.35cm}
Clean Ag(111) was prepared by cycles that consisted of 30\,min Ar-ion sputtering at an energy of ${E_{\text{Ar}} = 500}$\,eV 
and consecutive annealing at ${T_{\text{ann}} \approx 700}$\,K for 20\,min. 
In order to achieve the well-ordered BiAg$_2$ alloy with the $(\sqrt{3} \times \sqrt{3})$Bi/Ag(111)$R30^{\circ}$ reconstruction, 
$1/3$ of a pseudomorphic monolayer Bi was deposited onto the clean Ag(111) surface from a home-built Knudsen cell evaporator. 
During the deposition of Bi, the sample was held at elevated temperatures of ${T_{\text{sample}} \approx 550}$\,K. 
To reduce the defect density, the sample was afterwards held at ${T_{\text{sample}} \approx 500}$\,K 
for one more minute \cite{Ast2007,ElKareh2013,ElKareh2014,Leisegang2023}. 
H$_2$Pc molecules (Sigma-Aldrich) were deposited from a four-pocket Knudsen cell evaporator (Dodecon) 
onto the sample held at room temperature \cite{Leisegang2023}.
\vspace{-0.3cm}

\subsection*{Molecule manipulation}
\label{subsec:manipulation}
\vspace{-0.35cm}
As reported previously \cite{Leisegang2023}, H$_2$Pc molecules 
tend to adsorb at step edges or defects rather than on flat terraces of the BiAg$_2$ surfaces. 
Therefore, single molecules had to be moved to a defect-free surface area by means of STM manipulation. 
The manipulation was performed while scanning over the molecule and thereby dragging it.
Typical tunneling parameters for this process were ${U_{\text{bias}} \leq 20}$\,mV and ${I_{\text{set}} > 5}$\,nA. 
Eventually, the excitation barrier of the detector molecule was reduced 
by deprotonation of H$_2$Pc to HPc with a voltage pulse ${U_{\text{bias}} \geq 2.5}$\,V.
\vspace{-0.35cm}

\subsection*{Tip preparation}
\label{subsec:magnteic_tip}
\vspace{-0.35cm}
In order to obtain magnetically sensitive tips we used the procedure described previously \cite{Haertl2022}. 
In short, the freshly etched W tips was flash-heated under UHV conditions 
and then dipped several nanometers into a $200$\,AL thick Gd film on a W(110) substrate. 
Occasionally, a gentle voltage pulse of ${U_{\text{bias}} \geq \pm 4}$\,V was applied between the Gd surface and the tip. 
A more detailed explanation on this tip preparation is given in Suppl.\ Sect.~5.
\vspace{-0.3cm}

\subsection*{Tip characterization}
\label{subsec:tip_characterization}
\vspace{-0.35cm}
To prepare unpolarized tips, a W tip was flash-heated under UHV conditions to remove any possible contamination with magnetic material. 
Each magnetically sensitive Gd-coated tip was characterized before and after utilizing them for MONA. 
Before the transport measurements, the magnetic sensitivity of the STM tip was verified 
by imaging the magnetic domain structure of $200$\,AL thick Gd films on W(110) \cite{Haertl2022}, also see Supplementary Section 5. 
To unambiguously prove the existence of an in-plane component of the tip magnetization during MONA measurements, 
subsequent experiments were performed with the very same tip on Fe monolayer (ML) islands on W(110). 
This sample system is an ideal candidate for the post characterization since the Fe islands 
exhibit an in-plane magnetization pointing along the substrate's $\left[1\bar{1}0\right]$ direction \cite{Krause2007}. 
It was made sure that the in-plane contrast inverted upon a field sweep between $\pm 2$\,T, 
as confirmed by d$I$/d$U$ maps and spectroscopic data. 
For a detailed description and spin-polarized data on both substrates, see Suppl.\ Sect.~6.
\vspace{-0.3cm}

\subsection*{The MONA technique}
\label{subsec:MONA_technique}
\vspace{-0.35cm}
With the novel molecular nanoprobe technique (MONA) it is possible to investigate ballistic transport on the nanometer scale. \cite{Leisegang2021,Kuegel2017,Leisegang2018,Leisegang2018a,Kuegel2018,Kuegel2019,Leisegang2020,Leisegang2023}. 
Hereby, reversible switching events of a single molecule (rotation and/or tautomerization) 
serve as a measure for transport of remotely induced hot charge carriers. 
In order to inject charge carriers and detect the state of the molecule by the very same STM tip, 
the following measurement protocol is used:  
(i) The initial state of the molecule is determined by a scan 
at non-invasive parameters ($U_{\text{scan}} = 100 \dots 200$\,mV, $I_{\text{scan}} = 30 \dots 40$\,pA); 
(ii) The STM tip is moved to the injection point at a distance $d$ away from the molecule 
where charge carriers are induced for a duration $t_{\text{exc}}$ with $U_{\text{exc}}, I_{\text{exc}}$; 
(iii) Subsequently, the final state of the molecule is recorded by a topographic scan (see (i)). 
To account for the statistical nature of this process and to reduce the standard deviation, 
we repeated this procedure up to 4000 times for each data point. 
All data are presented as electron yield, which result from dividing the number of observed tautomerization events $S$ 
by the amount of injected charge carriers $\left(I_{\text{exc}}\cdot t_{\text{exc}}\right)$. 
The electron yield is normalized in the form that all three occurring rotations of the HPc molecule on the BiAg$_2$ surface 
are considered equal and thus all three rotations are weighted by a factor of $1/3$ each to the total electron yield. 
A detailed rotation-resolved analysis can be found in the Supplementary for the respective measurements in sections 2-4.
The error of the electron yield $\eta$ can be calculated by the standard deviation of the measured tautomerization events $\sigma_{S}$, 
since the uncertainties of the current $I$ and the injection time $t$ are negligible compared to the error of the tautomerization events. 
The error for the asymmetry $A$ is obtained by gaussian error propagation.

\subsection*{The overall measurement procedure}
\label{subsec:meas_proc}
\vspace{-0.35cm}
The overall measurement procedure contains eight steps: 
(i) In-situ preparation and pre-characterization of the magnetic tip on Gd(0001)/W(110) films; 
(ii) Sample exchange to H$_2$Pc/BiAg$_2$; 
(iii) Manipulation of H$_2$Pc molecules from step edges into defect-free surface areas; 
(iv) Application of an out-of-plane magnetic field ($\pm 2$\,T) to align the tip; 
(v) MONA measurements in remanent field; 
(vi) Apply magnetic field in the opposite field direction to invert the tip polarization; 
(vii) MONA measurements in remanent field; 
(viii) Post-characterization of the magnetic tip on Fe/W(110) monolayer islands 
to verify the in-plane polarization and the inversion of the tip magnetization at $\pm 2$\,T.


\end{document}




\title{Supplementary Information for: \\
``Probing spin-dependent charge transport on the single nanometer scale''}

\author{Patrick H\"{a}rtl}
	\email[corresponding author: ]{patrick.haertl@physik.uni-wuerzburg.de}
	\address{Physikalisches Institut, Experimentelle Physik II, 
	Universit\"{a}t W\"{u}rzburg, Am Hubland, 97074 W\"{u}rzburg, Germany}	
\author{Markus Leisegang}
	\affiliation{Physikalisches Institut, Experimentelle Physik II, 
	Universit\"{a}t W\"{u}rzburg, Am Hubland, 97074 W\"{u}rzburg, Germany} 
\author{Jens K\"{u}gel}
	\affiliation{Physikalisches Institut, Experimentelle Physik II, 
	Universit\"{a}t W\"{u}rzburg, Am Hubland, 97074 W\"{u}rzburg, Germany} 	
\author{Matthias Bode} 
	\address{Physikalisches Institut, Experimentelle Physik II, 
	Universit\"{a}t W\"{u}rzburg, Am Hubland, 97074 W\"{u}rzburg, Germany}	
	\address{Wilhelm Conrad R{\"o}ntgen-Center for Complex Material Systems (RCCM), 
	Universit\"{a}t W\"{u}rzburg, Am Hubland, 97074 W\"{u}rzburg, Germany}    

\date{\today}
\pacs{}
\maketitle

\vspace{-3cm}
\section{I. Rotational states of HPc on BiAg$_2$}
\begin{figure}[b]
	\includegraphics[width=0.8\linewidth]{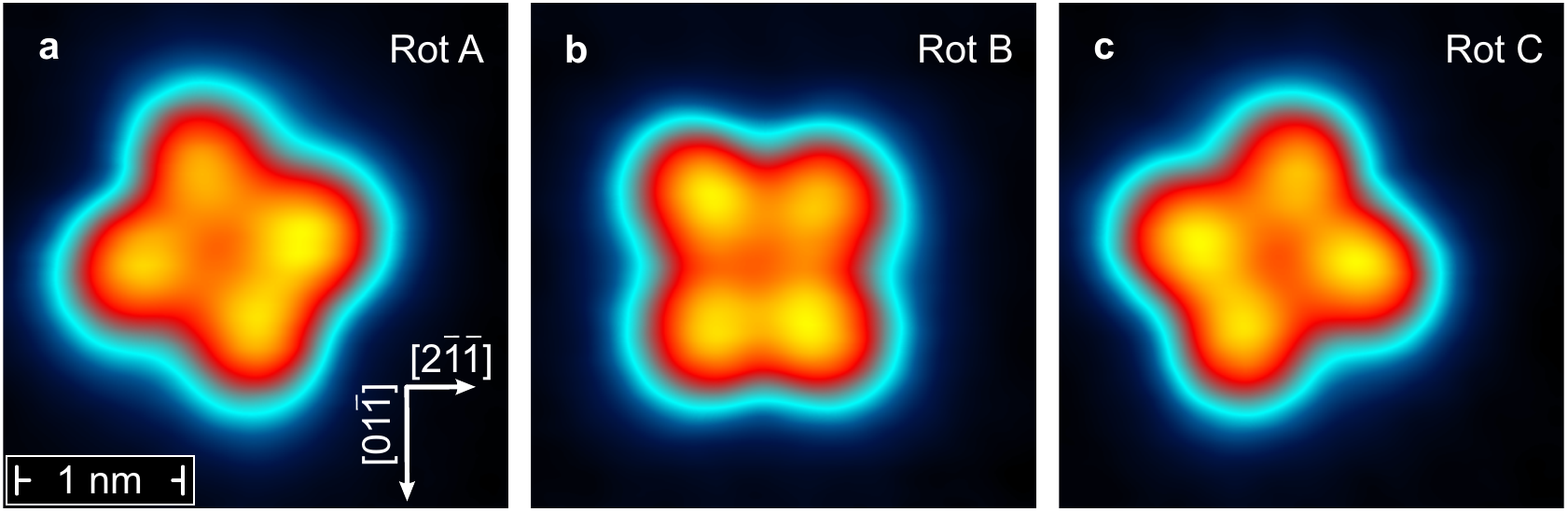}%
		\caption{\textbf{The three rotational states of HPC on BiAg$_2$.} 
		STM images of the states \textbf{a} Rot A, \textbf{b} Rot B, and \textbf{c} Rot C, 
		which are rotated by $120^{\circ}$ against each other. 
		Scan parameters: $U_{\text{scan}} = 100 \dots 200$\,meV, $I_{\text{scan}} = 30 \dots 40$\,pA.}
	\label{Fig:0} 
\end{figure}

HPc adsorbed on BiAg$_2$ exhibits, next to the four tautomeric states, three rotational states. 
As shown in the topographic images of Fig.\,\ref{Fig:0}, three rotational states of molecules coexist, 
which are rotated by $120^{\circ}$ towards each other and which will be labeled Rot A, Rot B, and Rot C hereafter. 
In each case the molecular arms are rotated by $45^{\circ}$ from the substrates high-symmetry axes. 
A detailed study of the rotational behavior and the influence of the individual states on remote excitation 
has been reported in a recent study by us \citep{Leisegang2022}. 
In this work we have shown that two distinct axes with different electron yields for tautomerization exist,
resulting in a significant influence of the rotational state on the remote excitation. 
Moreover, the MONA measurements have revealed that these three states Rot A, Rot B, and Rot C are equally populated, indicating their equivalence.

Within a MONA measurement, each data point includes several thousand single injection and detection events. 
Based on the statistical nature and the rather low probability for a rotation, however, 
which is by a factor of 30 lower as compared to a tautomerization \citep{Leisegang2022}, 
the absolute number of states Rot A, Rot B, and Rot C within an individual data set deviates. 
To account for this statistical variation and to correct the influence of an unequal state population, 
we weighted the experimentally determined tautomeric electron yield of all three rotational states by a factor of $1/3$ 
which results in a realistic expectation value.

In the following three sections, we will discuss the data presented in the main text with regard to the three rotational states. 
All data were recorded in remanence while the tip was pre-treated in an external out-of-plane oriented magnetic field of $\pm2$\,T.
The asymmetry $A_{\alpha}$ of the electron yields $\eta$ at a given angle $\alpha$ can be calculated as 
$A_{\alpha} = \left(\eta_{\alpha}^{+2\,\text{T}}-\eta_{\alpha}^{-2\,\text{T}}\right)/\left(\eta_{\alpha}^{+2\,\text{T}}+\eta_{\alpha}^{-2\,\text{T}} \right)$.

\section{II. Non-magnetic tip}
\begin{figure}[b]
	\begin{minipage}[b]{0.65\textwidth}
		\includegraphics[width=\linewidth]{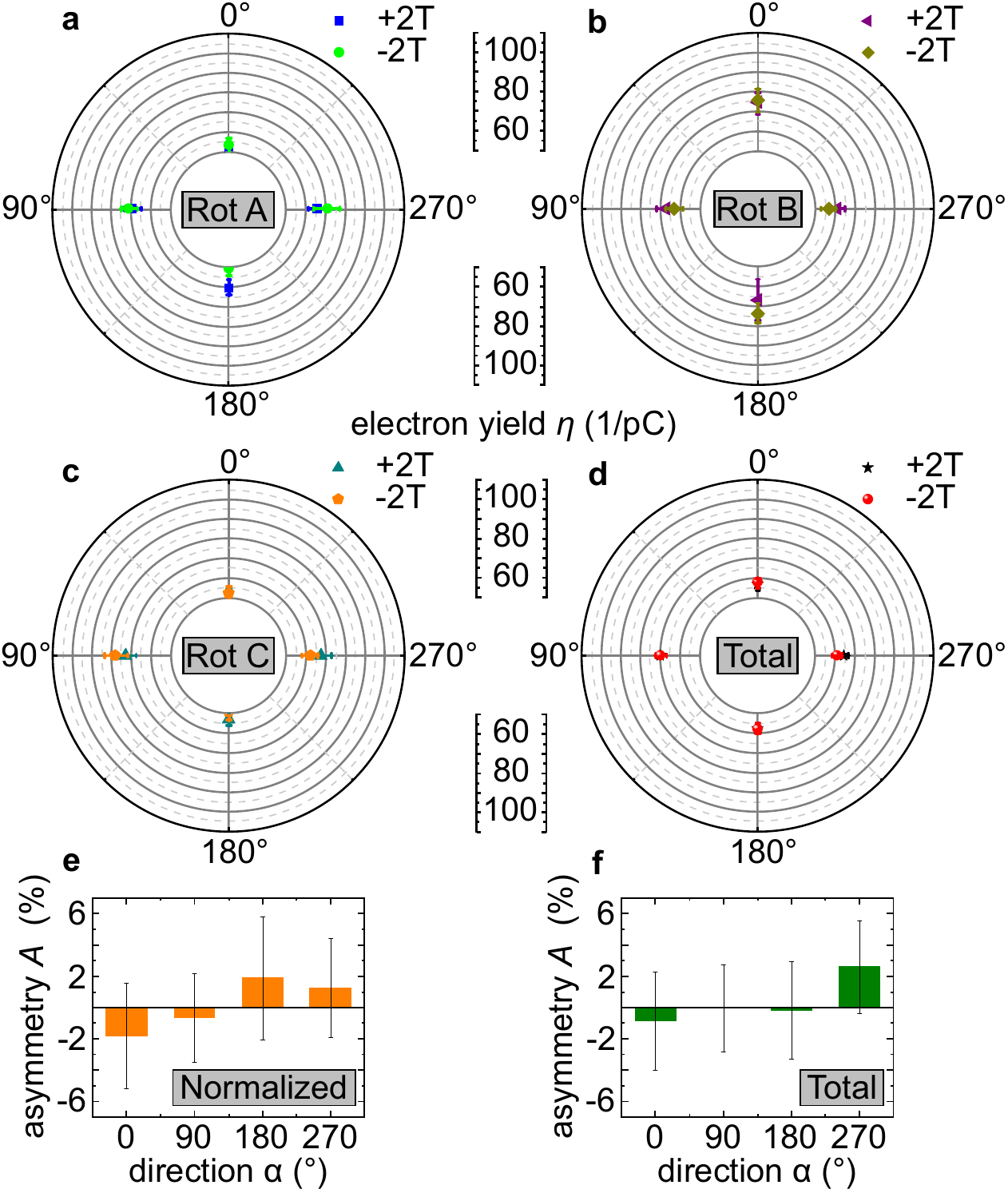}%
	\end{minipage}
	\hfill
	\begin{minipage}[b]{0.32\textwidth}
		\caption{\textbf{MONA data for an unpolarized W-tip.} 
		In \textbf{a}, \textbf{b}, and \textbf{c}, the data for the spin-averaged measurement set [main text Fig.\,2(a)] 
		are plotted individually for each molecular rotation Rot A, Rot B and Rot C, respectively. 
		All data have been measured in remanence (0\,T) after exposing the tip to an external magnetic field of $\pm2$\,T. 
		\textbf{d}, The non-weighted total electron yield, neglecting the influence of the rotations. 
		\textbf{e}, Weighted asymmetries of the data presented in Fig.\,2(a) in the main text. 
		\textbf{f}, Asymmetries of the non-weighted total electron yield presented in \textbf{d}.}
	\label{Fig:1} \vspace{1cm}
	\end{minipage}
\end{figure}

Figure~\ref{Fig:1} shows spin-averaged MONA data recorded with a non-magnetic tungsten tip 
for all three rotations individually (a-c) as well as the total electron yield in (d) (non-weighted). 
MONA parameters are $E_{\text{exc}} = 650$\,meV, $t_{\text{exc}} = 2.0$\,s, $I_{\text{exc}} = 1.0$\,nA. 
The corresponding weighted data are shown in Fig.\,2(a) of the main text. 
In all four plots, within the error bar the electron yield shows no correlation with the magnetic history, i.e., 
whether the data were recorded after a field sweep to positive or negative field.  

However, a significant deviation between different angles can be found within one rotational state 
as well as for the same angle between the three rotational states. 
While the $0^{\circ}-180^{\circ}$ axis shows a low electron yield for Rot A and Rot C, a high yield is observed along this axis for Rot B. 
This behavior inverts for the $90^{\circ}-270^{\circ}$ axis. 
Now Rot A and Rot C have a high and Rot B has a low electron yield. 
As reported in detail in our recent publication, these observations can be explained by the combination of the four fold symmetric molecular cage with the six fold symmetric substrate \citep{Leisegang2022}. 
Since two rotational states (Rot A and Rot B) exhibit the same behavior, 
they dominate the total electron yield (non-weighted) shown in Fig.~\ref{Fig:1}(d) 
where the $90^{\circ}-270^{\circ}$ axis has a higher yield as compared to the $0^{\circ}-180^{\circ}$. 
For comparison, the calculated asymmetry $A$ for the weighted and not-weighted averaged electron yield 
are presented in Fig.~\ref{Fig:1}(e) and (d), respectively. 
In both cases, within the error bars no significant asymmetry can be found.

\section{III. Magnetic tip}
\begin{figure}[t]
	\begin{minipage}[b]{0.65\textwidth}
		\includegraphics[width=\linewidth]{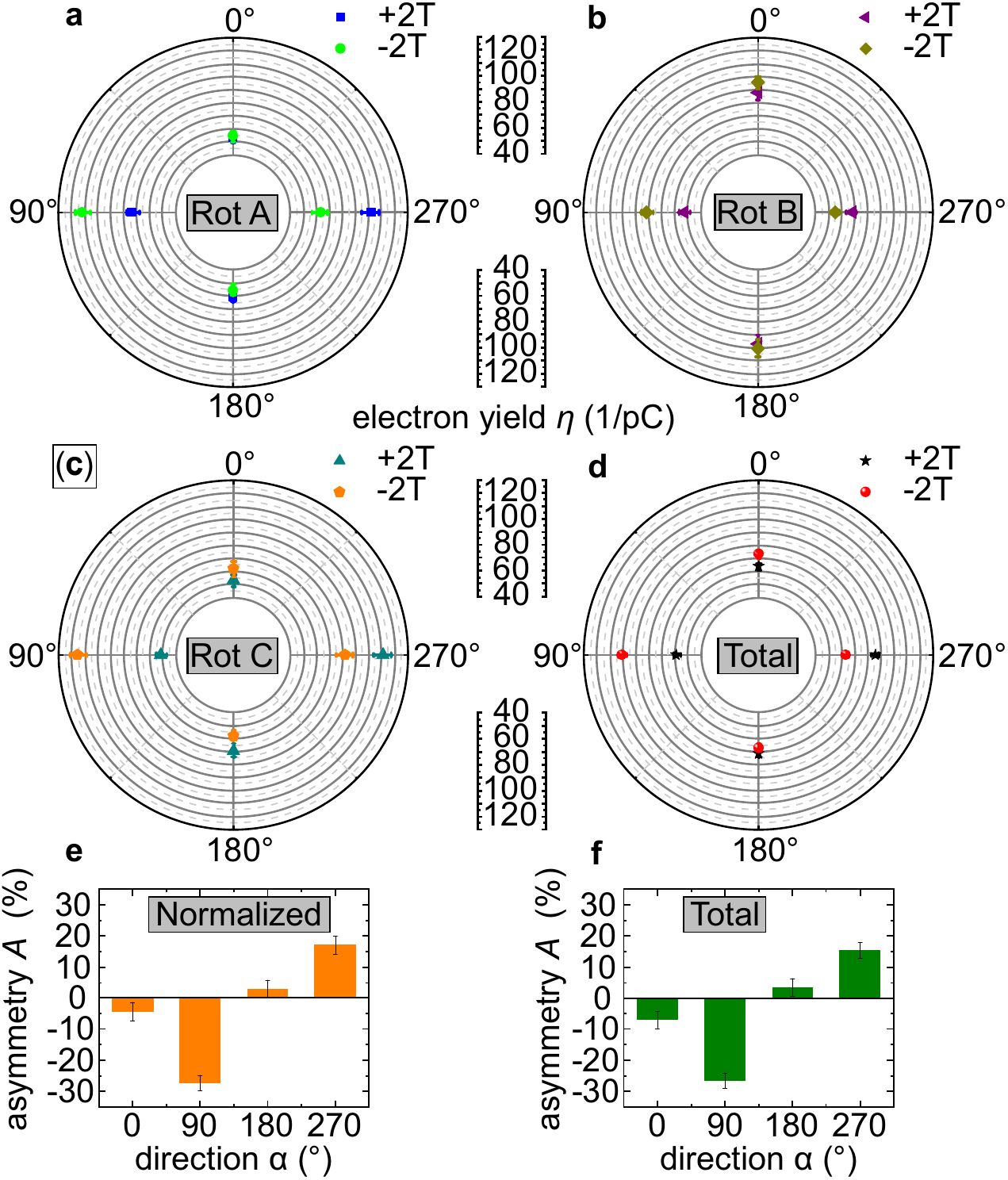}%
	\end{minipage}
	\hfill
	\begin{minipage}[b]{0.32\textwidth}
		\caption{\textbf{MONA data measured with a magnetically coated Gd tip.} 
		In \textbf{a}, \textbf{b} and \textbf{c} the data for the spin-polarized measurement set [main text Fig.\,2(b)] 
		are plotted individually for each molecular rotation Rot A, Rot B and Rot C, respectively. 
		All data have been measured in remanence (0\,T) after exposing the tip to an external magnetic field of $\pm2$\,T. 
		\textbf{d}, The non-weighted total electron yield, neglecting the influence of the rotations. 
		\textbf{e}, Weighted asymmetries for the data presented in Fig.\,2(b) in the main text. 
		\textbf{f}, Asymmetries for the non-weighted total electron yield presented in \textbf{d}.}
	\label{Fig:2} \vspace{1cm}
	\end{minipage}
\end{figure}

Figure~\ref{Fig:2} shows spin-polarized MONA data recorded with a magnetically Gd-coated tip 
for the three rotational states (a-c), as well as the total electron yield in (d) (non-weighted). 
The corresponding weighted data are shown in Fig.\,2(b) of the main text. 
MONA parameters are $E_{\text{exc}} = 650$\,meV, $t_{\text{exc}} = 2.0$\,s, 
$I_{\text{exc}} = 1.0$\,nA at a distance of $d = 4.0$\,nm from the molecule under four different angles.
All four plots show a qualitative similar behavior. 
The two-fold symmetry of the electron yield with an inversion for Rot B 
is equivalent to what has been discussed for the unpolarized measurements. 
However, a significant deviation between the field sweeps can be observed.
While the electron yields for both field sweeps ($\pm 2$\,T) are comparable along the $0^{\circ}-180^{\circ}$ direction, 
a significant difference can be observed along the $90^{\circ}-270^{\circ}$ axis for all three rotations as well as for the total sum. 
The calculated asymmetries $A$ for the weighted electron yields of the main text (Fig.\,2(b)) 
as well as for the non-weighted total yield of Fig.~\ref{Fig:2} (d) are plotted in Fig.~\ref{Fig:2}(e) and (f), respectively. 
Both plots reveal a similar behavior of $A$, while the absolute values 
slightly differ between the weighted and non-weighted electron yield $\eta$. 
Since all three rotations exhibit a significant asymmetry along the $90^{\circ}-270^{\circ}$ axis, 
we can exclude a direct influence by the rotations. 

\begin{figure}[t]
	\begin{minipage}[b]{0.65\textwidth}
		\includegraphics[width=\linewidth]{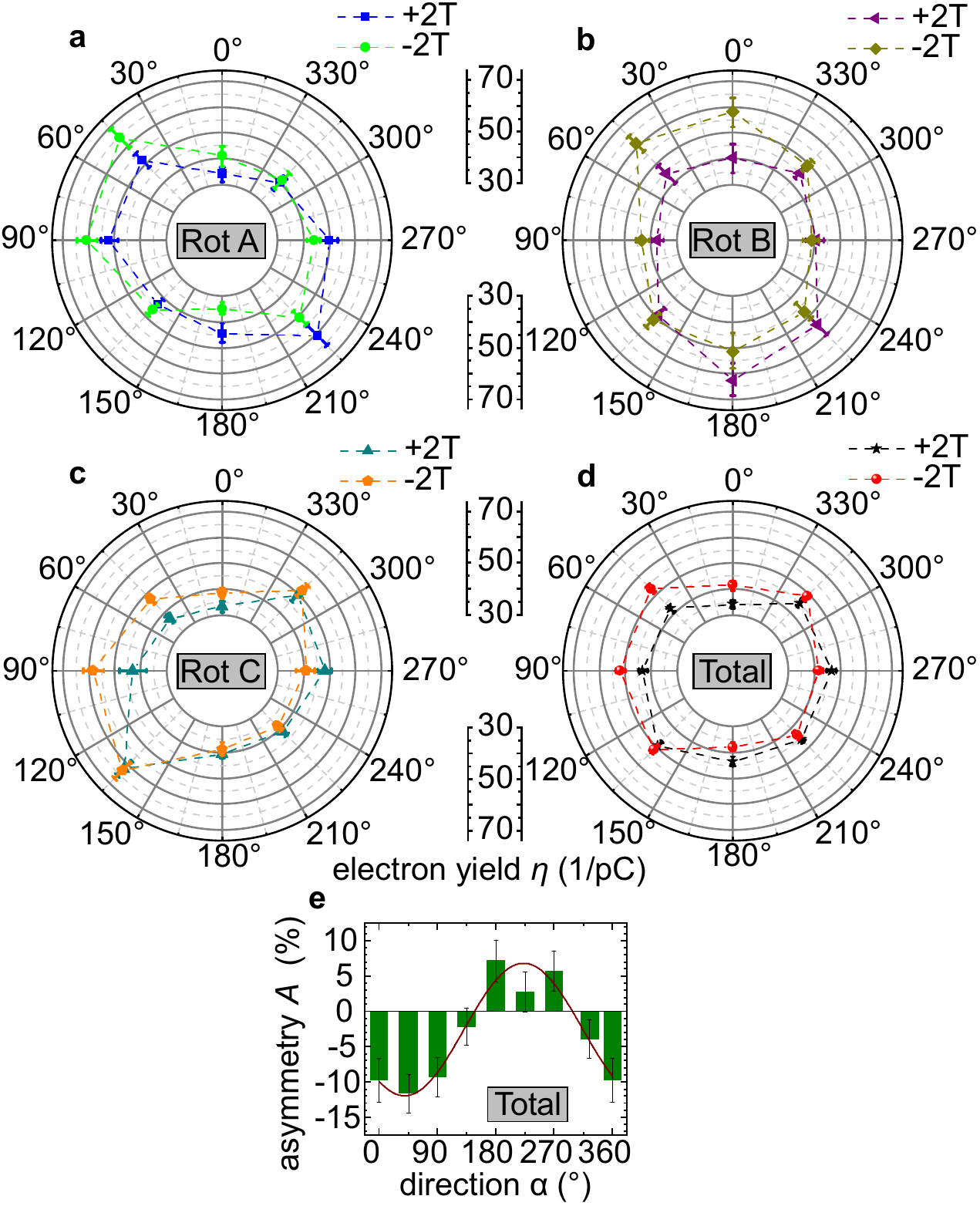}%
	\end{minipage}
	\hfill
	\begin{minipage}[b]{0.32\textwidth}
		\caption{\textbf{High angular resolution spin-polarized MONA measurement.} 
		In \textbf{a}, \textbf{b} and \textbf{c} the data for the spin-polarized measurement set [main text Fig.\,2(c)] 
		are plotted individually for each molecular rotation Rot A, Rot B and Rot C, respectively. 
		All data have been measured in remanence (0\,T) after exposing the tip to an external magnetic field of $\pm2$\,T. 		
		\textbf{d}, The non-weighted total electron yield, neglecting the influence of the rotations. 
		\textbf{e}, Asymmetries for the non-weighted total electron yield presented in \textbf{d} with a cosine fit (brown line). 
		The dashed lines in \textbf{a}, \textbf{b}, \textbf{c}, and \textbf{d} serve as a guide to the eye only.}
	\label{Fig:3} \vspace{1cm}
	\end{minipage}
\end{figure}

Additional SP-MONA data, acquired with a macroscopically different magnetically Gd-coated tip, are presented in Figure~\ref{Fig:3}. 
MONA parameters are $E_{\text{exc}} = 650$\,meV, $t_{\text{exc}} = 2.0$\,s, 
$I_{\text{exc}} = 1.5$\,nA at a distance of $d = 4.5$\,nm from the molecule. 
These measurements provide a higher angular resolution with eight angles 
which are again plotted separately for all three rotations as well as the non-weighted total yield. 
The corresponding weighted data are shown in the main text Fig.\,2(c). The dashed lines serve only as a guide to the eye.
Due to the higher resolution, a two-fold mirror symmetric electron yield can be observed for all three rotations. 
The plots show a rotation of roughly $120^{\circ}$ towards each other, which is in line with the molecular rotations on the surface. 
However, the axis of high and low asymmetry $A$ are always pointing 
along the $45^{\circ}-225^{\circ}$ and $135^{\circ}-315^{\circ}$ axes, respectively. 
These axes can also be identified in the non-weighted total electron yield in Fig.~\ref{Fig:3}(d). 
The resulting asymmetry is shown in Figure~\ref{Fig:3}(e), where a cosine-like behavior 
can be fitted to the graph in accordance to the weighted data shown in the main text Fig.\,2(c) and (d).

\vfill\newpage

\section{IV. MONA on a cluster}
\begin{figure}[t]
	\begin{minipage}[b]{0.65\textwidth}
		\includegraphics[width=\linewidth]{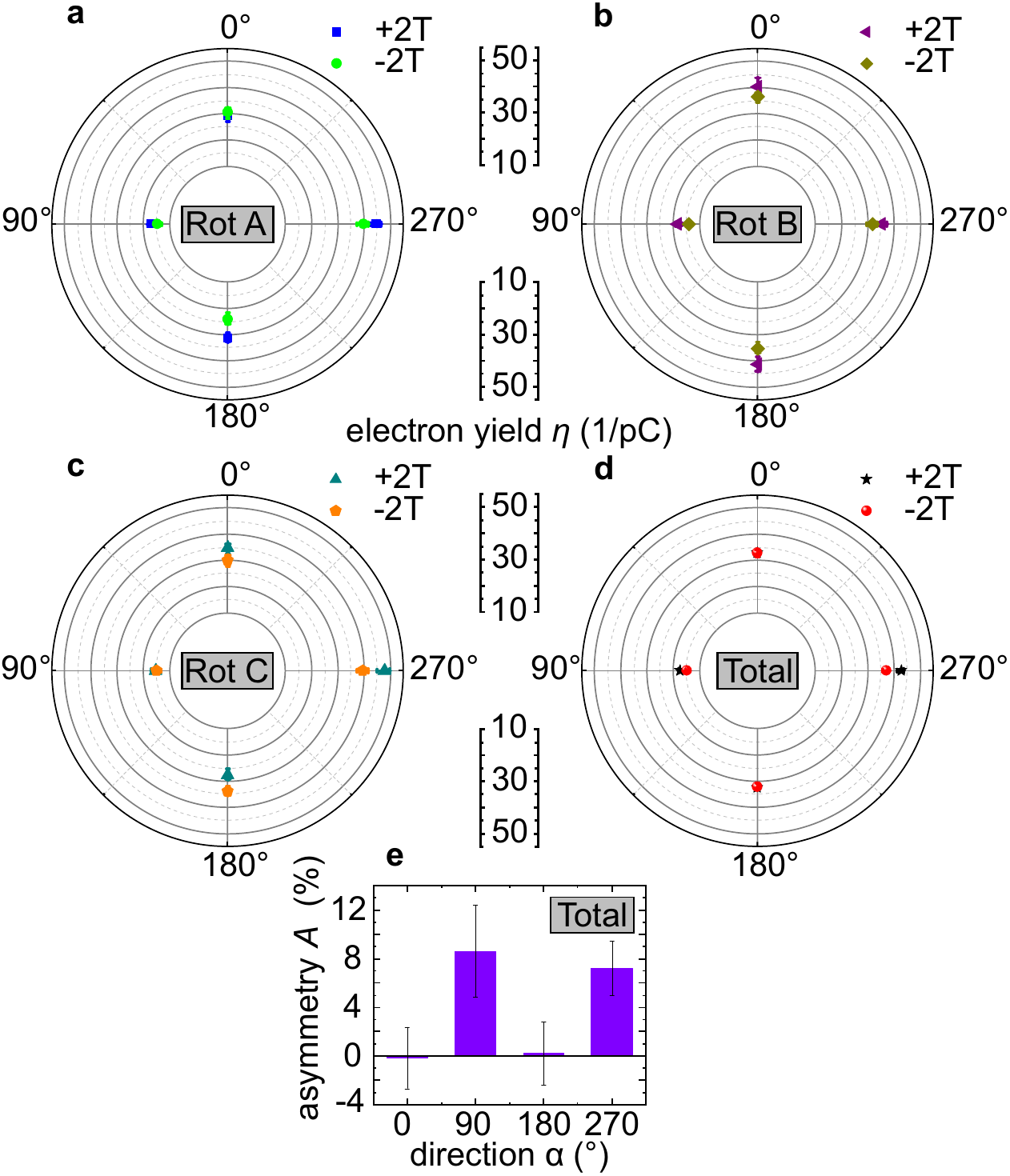}%
	\end{minipage}
	\hfill
	\begin{minipage}[b]{0.32\textwidth}
		\caption{\textbf{SP-MONA data with a Gd-cluster.} 
		In \textbf{a}, \textbf{b} and \textbf{c} the data for the spin-polarized measurement set [main text Fig.\,3(b)] 
		are plotted individually for each molecular rotation Rot A, Rot B and Rot C, respectively. 
		All data have been measured in remanence (0\,T) after exposing the tip to an external magnetic field of $\pm2$\,T. 	
		\textbf{d}, The non-weighted total electron yield, neglecting the influence of the rotations. 
		\textbf{e}, Asymmetries for the non-weighted total electron yield presented in \textbf{d}.}
	\label{Fig:4} \vspace{1cm}
	\end{minipage}
\end{figure}

In this measurement, a Gd-cluster has been placed in the $90^{\circ}$ direction as a magnetic impurity.
Charge carriers were injected by a Gd-coated tip. 
MONA parameters are $E_{\text{exc}} = 650$\,meV, $t_{\text{exc}} = 2.5$\,s, and $I_{\text{exc}} = 2.0$\,nA 
at a distance of $d = 6.5$\,nm from the molecule.
The data are presented equivalently to the spin-averaged 
and spin-polarized data sets of Figs.~\ref{Fig:1} and \ref{Fig:2}, respectively. 
In the three plots of Fig.~\ref{Fig:4}(a)-(c) we document the results obtained for all molecular rotation listed above.  
The averaged data of Fig.~\ref{Fig:4}(d) clearly reveal the influence of the Gd cluster under $90^{\circ}$. 
The asymmetry shown in Fig.~\ref{Fig:4}(e) clearly deviates from a cosine-like behavior, 
highlighting the strong influence of the cluster. 
The weighted data of this experimental run are shown in the main text Fig.~3(b) and (d).
\vfill\newpage

\begin{figure}[t]
	\begin{minipage}[b]{0.65\textwidth}
		\includegraphics[width=\linewidth]{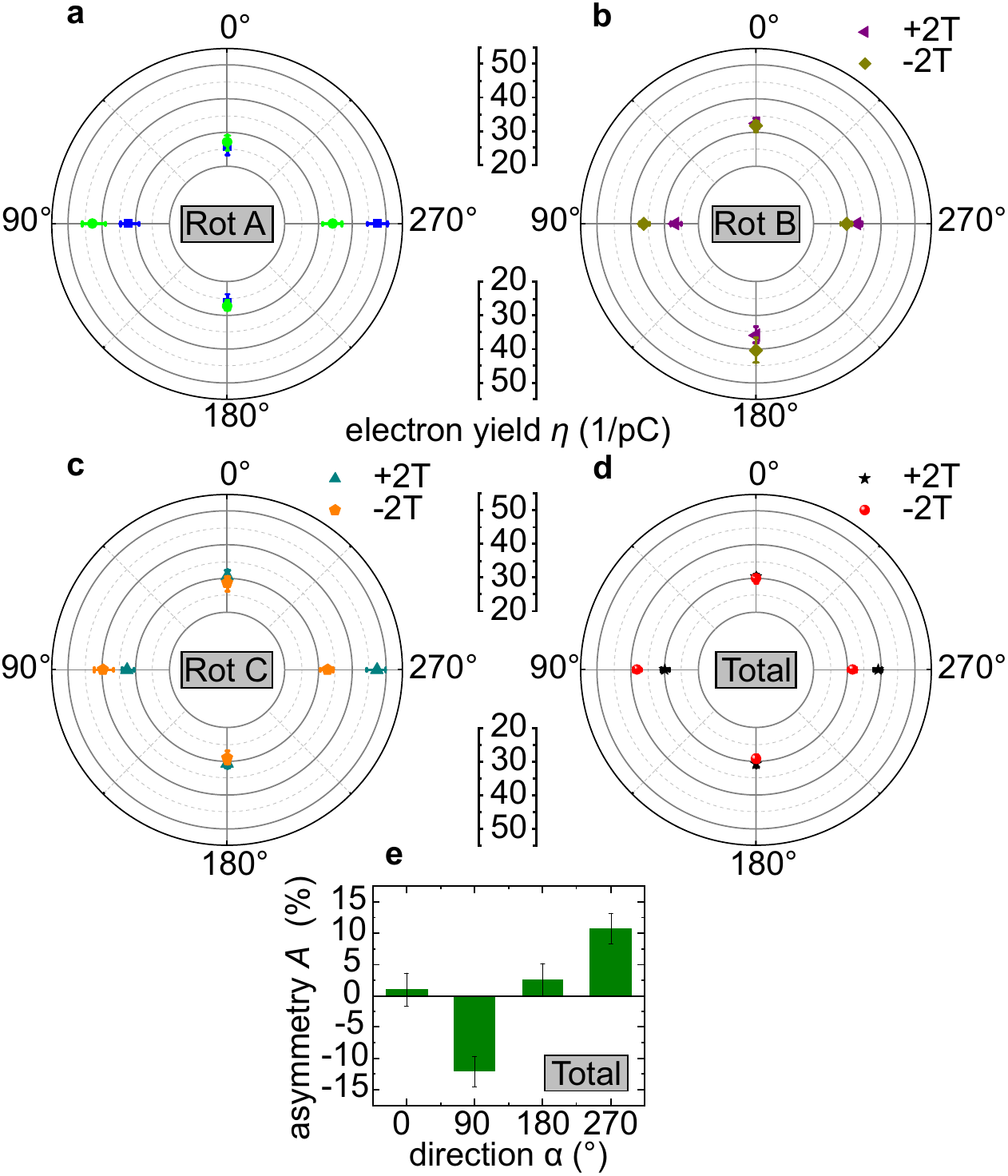}%
	\end{minipage}
	\hfill
	\begin{minipage}[b]{0.32\textwidth}
		\caption{\textbf{Reference SP-MONA data without a Gd-cluster.} 
		In \textbf{a}, \textbf{b} and \textbf{c} the data for the spin-polarized measurement set [main text Fig.\,3(c)] 
		are plotted individually for each molecular rotation Rot A, Rot B and Rot C, respectively. 
		All data have been measured in remanence (0\,T) after exposing the tip to an external magnetic field of $\pm2$\,T. 	
		\textbf{d}, The non-weighted total electron yield, neglecting the influence of the rotations. 
		\textbf{e}, Asymmetries for the non-weighted total electron yield presented in \textbf{d}.}
	\label{Fig:5} \vspace{1cm}
	\end{minipage}
\end{figure}

An equivalent representation is given for a HPc without any defect nearby, acting as reference system for the measurements with a cluster. This setup has been probed on the very same surface with the same Gd-coated tip in a distance of around 30\,nm to the setup with cluster.
The data in Figure~\ref{Fig:5} show an equivalent cosine-like behavior (see asymmetry in Figure~\ref{Fig:5}(e)) as discussed for the SP-MONA data in Figure~\ref{Fig:2} as well as the same signature originating from the molecular rotations.

\section{V. Pre-characterization of the magnetic tip on Gd/W(110)}
\phantomsection \label{sec:Gd}
In this section, the preparation and characterization of a magnetically Gd-coated tip is described. 
W tips are electro-chemically etched in a NaOH solution, carefully rinsed with distilled water, 
transferred into the ultra-high vacuum (UHV) system via a load lock, 
and then cleaned in several high-temperature flashes from potential contamination by e-beam heating.  

\begin{figure}[t]
	\includegraphics[width=1.0\linewidth]{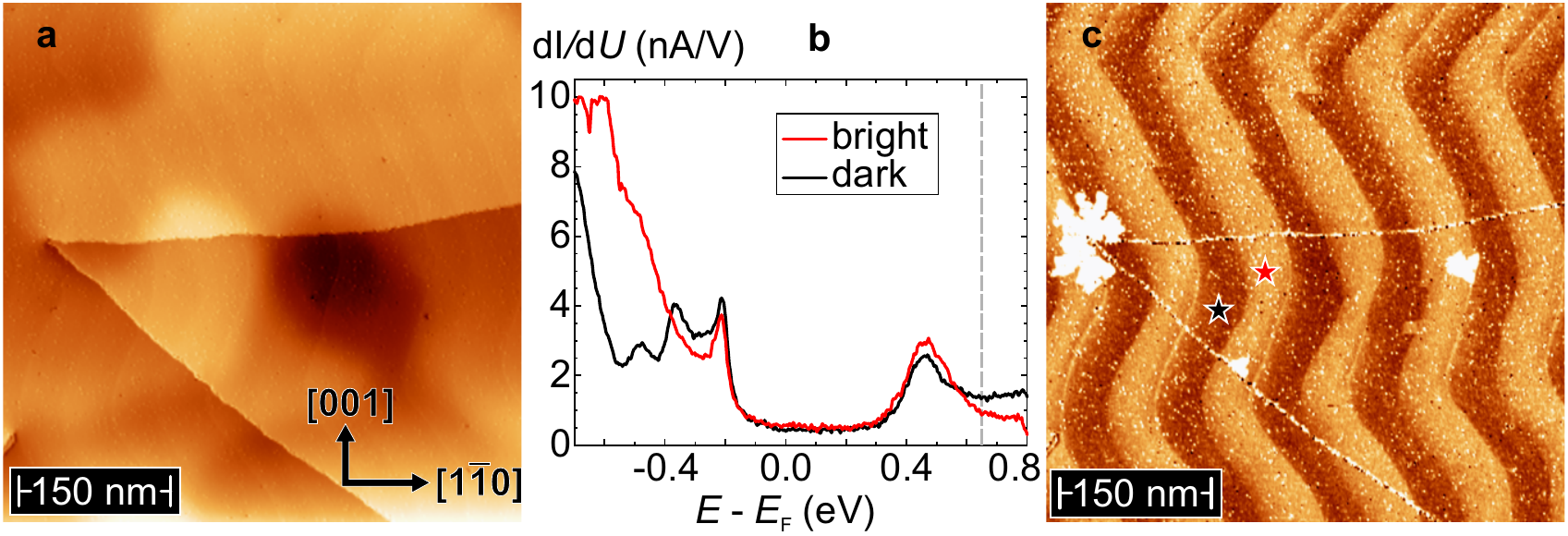}%
		\caption{\textbf{Tip pre-characterization on Gd/W(110).} 
		\textbf{a}, Overview scan of a 200\,AL Gd film epitaxially grown on a W(110) substrate. 
		\textbf{b}, d$I$/d$U$ point spectra taken at the positions marked by a red/black star in (\textbf{c}), 
		revealing a tip polarization at $E - E_{\text{F}} = E_{\text{exc}} = 650$\,meV (grey dashed line). 
		\textbf{c}, Magnetically sensitive d$I$/d$U$ map of the film presented in (\textbf{a}). 
		A zig-zag like domain pattern along the W[001] direction can be identified, 
		comparable to the data already published by H\"{a}rtl \emph{et al.} \cite{Haertl2022}. 
		Scan parameters: $U_{\text{bias}} = 650$\,mV, $I_{\text{set}} = 1$\,nA.}
	\label{Fig:6} 
\end{figure}

To magnetize these initially non-magnetic tips, we utilize Gd(0001)/W(110) films 
which were investigated by us in a recent SP-STM study \cite{Haertl2022}. 
Fig~\ref{Fig:6}(a) shows the topography of a 200\,AL Gd film grown on a W(110) substrate. 
This film was deposited at a deposition rate of $(19.8\pm 3.7)$\,AL per minute onto the substrate held at room temperature. 
Subsequent annealing for five minutes at $T_{\text{anneal}} \geq 900$\,K results 
in a very flat and highly ordered surface, see Ref.~\onlinecite{Haertl2022} for details. 

The transfer of magnetic material (Gd) from this sample to the tip is achieved 
by dipping the tip a few nanometers into the Gd film or by gentle pulsing $U_{\text{bias}} > 4$\,V.  
To our experience both procedures may result in the attachment of a Gd cluster to the tip. 
To verify that the transfer resulted in a magnetically sensitive tip, 
we measured differential conductance d$I$/d$U$ maps at $E-E_{\text{F}} = E_{\text{exc}} = 650$\,meV. 
Only if the tip is indeed magnetically sensitive, 
a zig-zag pattern with two distinct contrast levels like the one shown in Fig.\,\ref{Fig:6}(c) appears, 
which is characteristic for the domain structure of Gd films in this thickness range \cite{Haertl2022}.  
At the points marked in the bright (red star) and the dark area (black star), 
d$I$/d$U$ point spectra were taken at a frequency of $f_{\text{mod}} = 5.309$\,kHz 
(well above the cutoff frequency of the feedback loop) and with a modulation of $U_{\text{bias}} = 10$\,mV. 
These spectra are plotted in Fig.\,\ref{Fig:6}(b).
The significant difference between theses spectra evidences its suitability for spin-polarized experiments. 
Most important for the measurements presented in this contribution is the fact 
that we find a significant polarization at $E-E_{\text{F}} = E_{\text{exc}} = 650$\,meV, marked by a grey dashed line in Fig.\,\ref{Fig:6}(b).

\section{VI. Post-characterization of the magnetic tip on Fe/W(110) ML-islands}
\label{sec:Fe}
After successful SP-MONA measurements on BiAg$_2$, 
we verified the tip polarization by post-characterization on Fe/W(110) monolayer islands. 
This is an ideal test system, since the islands exhibit a well-defined in-plane magnetization 
along the $\left[\bar{1}\bar{1}0\right]$ direction of the W(110) substrate \cite{Krause2007}. 
This direction is aligned with the $0^{\circ}-180^{\circ}$ axis in our SP-MONA experiments.

\begin{figure}[t]
	\includegraphics[width=1.0\linewidth]{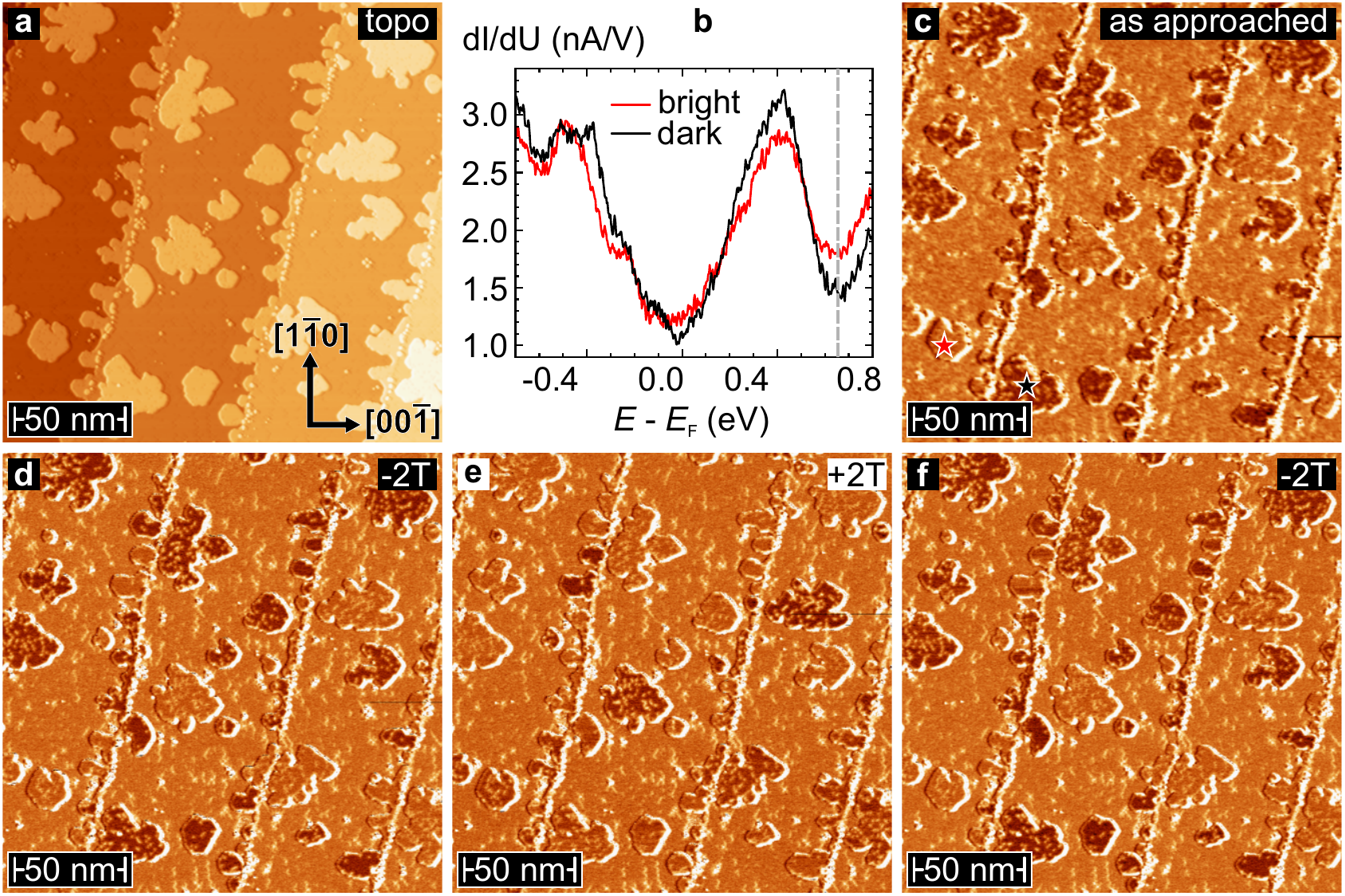}%
		\caption{\textbf{Post-characterization of the magnetic tip on Fe/W(110) ML-islands.} 
		\textbf{a}, Topography of Fe monolayer islands on W(110) at a Fe coverage of $\approx 0.24$\,AL. 
		\textbf{b} d$I$/d$U$ point spectra measured on the Fe islands marked with a red/black star in (\textbf{c}), 
		revealing a significant tip polarization at $E - E_{\text{F}}$ = 650\,meV (see grey hatched line) 
		after the SP-MONA measurements. 
		\textbf{c}, d$I$/d$U$ map directly recorded after the SP-MONA measurement presented in Fig.\,3 of the main text. 
		The islands show two different contrast levels, 
		indicating a significant in-plane polarization component of the magnetic STM tip. 
		\textbf{d}, d$I$/d$U$ map of the region shown in (\textbf{c}) after tip-treatment at -2\,T, measured in remanence. 
		\textbf{e}, d$I$/d$U$ map after a field ramp to +2\,T, measured in remanence. 
		\textbf{f}, d$I$/d$U$ map after a field ramp back to -2\,T, where the islands again inverted their contrast. 
		STM parameters: $U_{\text{bias}} = 650$\,mV, $I_{\text{set}} = 1$\,nA.}
	\label{Fig:7} 
\end{figure}

Fe was deposited onto the clean W(110) substrate from a rod with a diameter of 2\,mm 
via an electron beam evaporator at a flux of 10\,nA for 30\,s deposition, leading to a coverage of $\approx 0.24$\,AL. 
During growth the substrate is held at a temperature $T_{\text{sample}}\approx 400$\,K. 
The resulting islands on the stepped substrate can be seen in the overview scan of Fig.\ref{Fig:7}(a). 

To verify that the tip is sensitive to the in-plane component of the magnetization 
at the energy used for our SP-MONA experiments, i.e., $E - E_{\text{F}} = E_{\text{exc}} = 650$\,meV, 
we measured differential conductance d$I$/d$U$ maps at exactly this energy.  
The corresponding image in Fig.\,\ref{Fig:7}(c) shows islands with two different contrast levels, 
indicating a significant in-plane polarization of the tip after the SP-MONA measurements. 
At the positions marked with the red (bright island) star and the black (dark island) star, d$I$/d$U$ point spectra are taken. 
The resulting spectra are plotted in Fig.\,\ref{Fig:7}(b), 
confirming a significant spin polarization at $E - E_{\text{F}} = E_{\text{exc}} = 650$\,meV (grey dashed line). 

Subsequently, we verified that the in-plane component along the $\left[\bar{1}\bar{1}0\right]$ direction 
of the W(110) substrate can indeed be reversed successfully by the application of an external magnetic field.  
We exposed the tip to fields of $-2$\,T, $+2$\,T and back to $-2$\,T.  
After each field ramp the magnetic state of the tip was investigated at remanence (0\,T).
For this purpose, we measured d$I$/d$U$ images taken at the same sample location.  
As can be seen by comparing the data sets presented in Fig.\,\ref{Fig:7}(d)-(f), 
the contrast levels of the islands inverted after each field sweep, thereby verifying the fact 
that the magnetization of the spin-sensitive tip used for the SP-MONA measurements presented in Fig.\,3 
can be (i) reversed with the available external magnetic fields, 
(ii) exhibits a significant spin polarization at the energy used in SP-MONA experiments ($E - E_{\text{F}} = E_{\text{exc}} = 650$\,meV), 
and (iii) that the tip magnetization exhibits a significant in-plane component, 
i.e., in the plane where a spin-polarized transport of spin--momentum-locked Rashba states is expected.



%